\documentclass[10pt,journal]{IEEEtran}
\usepackage{subfigure}

\usepackage{gensymb}

\ifCLASSINFOpdf
  \usepackage[pdftex]{graphicx}
  \graphicspath{{/figures/}{../pdf/}{../jpeg/}}
  \DeclareGraphicsExtensions{.pdf,.jpeg,.png}
\else
   \usepackage[dvips]{graphicx}
  \graphicspath{{figures/}{../pdf/}{../jpeg/}}
  \DeclareGraphicsExtensions{.eps,.png}
\fi
\usepackage[cmex10]{amsmath}
\usepackage[]{algorithm}
\usepackage{algpseudocode}

\DeclareMathOperator*{\argmin}{\arg\!\min}

\DeclareMathOperator*{\rank}{rank}
\DeclareMathOperator*{\tr}{tr}

\usepackage{amssymb}

\begin{document}

%
\title{\huge Constant Modulus Beamforming via Convex Optimization}
%
%
%

\author{Amir Adler,~\IEEEmembership{Member,~IEEE,}
        and~Mati~Wax,~\IEEEmembership{Fellow,~IEEE}
\thanks{A. Adler    e-mail: adleram@mit.edu,  M. Wax e-mail: mati.wax@gmail.com. }
}
\maketitle
\begin{abstract}
We present novel convex-optimization-based solutions  to the problem of blind beamforming of constant modulus signals, and to the related problem of linearly constrained  blind beamforming of constant modulus signals. These solutions ensure global optimality and are parameter free, namely, do not contain any tuneable parameters and do not require any a-priori parameter settings. The performance of these solutions, as demonstrated by simulated data, is  superior to  existing methods.
\end{abstract}

\begin{IEEEkeywords}
constant modulus algorithm, linearly constrained constant modulus algorithm, trace norm.\end{IEEEkeywords}

%
\IEEEpeerreviewmaketitle

\section{Introduction}
%
%
%
%
 Constant modulus (CM) beamforming is a well-known \textit{blind} array processing technique,  based on exploiting the constant modulus of the desired signal. It was  introduced and developed  in [1]-[7], following  the  pioneering works of Godard [1] and Triechler et al. [2] on blind CM equalization and    further improved in [8]-[9]. An extensive review is presented in [10]. It was also extended in [11]-[13] to    allow    additional linear constraints to be imposed on  the beamforming vector.\indent In spite of these developments, 
the main difficulty in CM beamforming is still largely unsolved - being cast as a  multidimensional non-convex minimization problem with multiple local minima [14]-[15], making global minimization very challenging. { 

\indent
   In this letter we present a novel solution  to the CM problem, based on  convex optimization  formulation [16]-[21]. This solution assures \textit{global optimality} and is \textit{parameter free}, i.e, it does not contain any tuneable parameters and does not require any apriori parameter setting. The  solution is then readily extended  to   enable additional linear constrains on the beamforming vector, which if properly constructed, are shown to provide further performance improvements. 

The rest of the letter is organized as follows. The problem formulation is presented in section II.  Section III describes the convex optimization solution. Sections IV describes the extension of the solution to the case of  linearly constrained CM. The computation time and the performance  of the solution are presented in section V. Finally, section VI presents the  conclusions. \section{Problem Formulation}
Suppose we want to receive a constant modulus signal $s_0(t)$, using an antenna array composed of $P$ antennas with arbitrary locations and arbitrary directional characteristics. Assume that the desired signal is impinging on the array from an unknown direction-of-arrival $\theta_0$, and that   $Q-1$ other interfering signals  $s_q(t), q=1,\ldots,Q-1$, are   also impinging  on the array from  unknown directions-of-arrival $\theta_1,\ldots,\theta_{Q-1}$. 
All the signals  are assumed to be narrow-band, namely that the array aperture, denoted by $d$, obeys $d<<c/B$, where $c$ is the speed of light, and $B$ is the signals bandwidth.  
 
Under these assumptions, the $P\times 1$   vector $\mathbf x(t)$  of the  complex envelopes of  the signals received by the array can be written as:
\begin{equation}
\mathbf{x(t)} = \mathbf{a(\theta_0)s_0(t)} + \sum_{q=1}^{Q-1} {\mathbf{a}(\theta_q)s_q(t)} + \mathbf{n(t)},
\label{eq:snapshot}
\end{equation}
where $\boldsymbol a(\theta_0)$  is the  $P\times 1$   steering vector of the array toward the desired CM signal  $s_0(t)$,     $\mathbf{a}(\theta_ q)$  is the  $P\times 1$  steering vector of the array toward the interfering signal  $s_q(t)$, and $\mathbf{n(t)}$  is the $P \times 1$  noise vector. 
We further assume that the number of impinging signals obeys $Q\le P$ and that the  signals' steering vectors $\{\mathbf a (\theta _q)\}^{Q-1}_{q=0}$ are \textit{linearly independent.}

 The blind beamforming problem  can be formulated as follows: 
 Given the sampled array vectors $\{\mathbf x(t_{n})\}_{n=1,}^N$, find a  $P \times 1$ beamforming weight vector  $\mathbf w$,  such that  the beamformer output $y(t) =\mathbf w^H \mathbf x(t)$, where $H$  denotes the   conjugate transpose, \textit{provides a good estimate of the CM  signal} $s_0(t)$.

\section{Convex Constant Modulus Algorithm} 
    Assuming, without loss of generality, that the modulus of   the desired signal $s_0(t)$ is 1, the  common Constant Modulus Algorithm  (CMA) cost function for estimating the beamforming weight $\mathbf w$ is given by the sample-average of the deviation of the beamformer power output from (1): \begin{equation}
\label{eq:CM_criterion}
\mathbf{\hat  w} =\argmin_{\mathbf {w}} \frac{1}{N}  \sum_{n=1}^N  (|\mathbf w^H \mathbf x(t_{n})|^2-1)^2.
\end{equation}
This is a fourth order minimization problem in the vector $\mathbf w$, and as such does not admit a closed form solution.  Moreover, as shown in [14]-[15], it  is a non-convex problem (i.e. it has multiple local minima), making global minimization very challenging. We  next show how to reformulate the CMA as a convex optimization problem, which assures global optimality. 
First, we rewrite the beamformer power output, denoted by  z(t), as\footnote{We use the following properties of the trace operator $\tr()$: (i) cyclic shift: $\tr(ABCD)=\tr(BCDA)=\tr(CDAB)=\tr(DABC)$; and (ii) $\tr(a)=a$ for any scalar $a$.}:
\begin{eqnarray}
 z(t)&=&|\mathbf w^H \mathbf x(t)|^2   = \mathbf w^H \mathbf  x(t)\mathbf x(t)^H\mathbf w\\&=& \tr(\mathbf w \mathbf w^H \mathbf  x(t)\mathbf x(t)^H)= \tr(\mathbf W  \mathbf x(t)\mathbf x(t)^H),\nonumber
\end{eqnarray}
where $\tr() $ denotes the trace of the bracketed matrix and $\boldsymbol W$ denotes  the $P \times P$ positive semidefinite rank-1 matrix: 
\begin{equation}
\mathbf W = \mathbf w \mathbf w^H. 
\end{equation}
We can now rewrite (2) as
\begin{subequations}
\begin{equation}       
\mathbf{\hat{W}} =\argmin_{\mathbf {W}} \frac{1}{N} \sum_{n=1}^N  |z(t_{n})-1|^2, \end{equation}
subject to: 
\begin{equation}
z(t_{n})=\tr(\mathbf W \mathbf  x(t_{n})\mathbf x(t_{n})^H)\;\;\;\;\;n=1,...,N \end{equation} 
\begin{equation}
\mathbf{W}\succcurlyeq  0, 
\end{equation}
\begin{equation}
\rank \mathbf W = 1,
\end{equation}
\end{subequations}
where $\mathbf{W}\succcurlyeq  0$ denotes the positive semidefinite constraint. Note however, that since the rank constraint (5d) is not convex, the minimization problem  is  not convex. A commonly-used  convex relaxation  surrogate to
the rank-1 constraint is to minimize the trace norm (nuclear norm), defined as the sum of the singular values of the matrix [17]-[19]. Recalling that  $\mathbf W$ is  a positive semidefinite matrix, it follows that its  trace norm is given by $\tr(\mathbf {W})$. This implies that  we can  reformulate  the CM problem as the following \textit{convex optimization} problem: 
\begin{subequations}
\begin{equation}
\mathbf  {\mathbf{\hat  W}} =\argmin_{\mathbf{W}} \{( \frac{1}{N}\sum_{n=1}^N  |z(t_n)-1|^2 )+ \tr( \mathbf W )\}, 
\end{equation}
 \vskip -0.35cm subject to:  
\begin{equation}
\tr (\mathbf W \mathbf  x(t_{n})\mathbf x(t_{n})^H)=z(t_{n})\;\;\;\;\;n=1,...,N. \end{equation}
\begin{equation}
\mathbf{W}\succcurlyeq 0.
\end{equation}
\end{subequations}
Since (6) is a convex optimization problem, we can use any of the convex optimization solvers [16]-[22] to solve for  $\hat {\mathbf W.}$
 
\indent
With  $\hat {\mathbf W}$  at hand,  a straightforward way to estimate the beamforming vector $\mathbf w$ is  by the rank-1 approximation of  $\hat {\mathbf W}$:    
\begin{equation}
\hat {\mathbf W} \simeq\ \mathbf \lambda _{1}\mathbf v_1 \mathbf v_1 ^{H} ,\end{equation}
where $\mathbf \lambda _{1}$ denotes  the largest eigenvalue of $\hat {\mathbf W,}$ and $\mathbf v_1$ denotes the eigenvector of $\hat {\mathbf W}$  corresponding to $\lambda _{1}$. Using this rank-1 approximation, we estimate the beamforming vector $\mathbf w$ as:  
\begin{equation}
\hat {\mathbf  w} = \mathbf v_1 . 
\end{equation}

\section{Convex Linearly Constrained CMA}
    
In many scenarios involving CM beamforming, it may be desired to impose additional constraints  on the beamformer vector in the form of the following linear constraint:
\begin{equation}
\mathbf w^H \mathbf C = \mathbf v^H,
\end{equation}
where $\mathbf C$  is a  $P \times K$  known matrix and $\mathbf v$  is a  $K\times 1$  known vector.
This problem is referred to as the Linearly Constrained Constant Modulus Algorithm (LCCMA).
 
An example for such a constraint is the well-known "look direction" constraint: \begin{equation}
\mathbf w^H \mathbf a(\theta) = 1, 
\end{equation}
constraining $\mathbf w$ to have a unity gain in the direction  $\theta$. Another example  is the  constraint,
\begin{equation}
\mathbf w^H \mathbf B = \mathbf 0, 
\end{equation}  
constraining $\mathbf w$ to be orthogonal to the columns of   $\mathbf B$. One example for such a $\mathbf B$ is 
\begin{equation}
 \mathbf B= \mathbf a(\theta), 
\end{equation} 
 assuring deep "nulls" in the direction  $\theta$. This  may be desired, for example, in case a strong interference is known to be impinging from direction $\theta$ and the desire is to put a deep null in this direction.   Another example is  
\begin{equation}
\mathbf B=[\mathbf v_{Q+1},...,\mathbf v_P], 
\end{equation}
where $\mathbf v_i$ is the eigenvector of the array covariance matrix 
$\mathbf {\hat R}=\sum_{n=1}^N x(t_n)x(t_n)^H$ corresponding to the $i$-th eigenvalue. This constraints $\mathbf w$ to be orthogonal to the noise subspace, i.e., to be confined to the $Q$-dimensional signal subspace [23]. This low-dimensional confinement reduces the number of degrees-of-freedom of $\mathbf w$, thereby improving the solution performance, especially in challenging conditions such as small number of samples and low signal-to-noise ratio. 

To incorporate the linear constraint (9) into our convex CMA formulation, we first rewrite it as 
\begin{equation}
\mathbf w^H \mathbf c_k = v_k,\;\;\;\;\;k=1,...,K 
\end{equation}
where $\mathbf c_k$ denotes the $k$-th column of  $\mathbf C$ and  $v_k$ denotes the  $k$-th element of $\mathbf v$.
Now, using the properties of the trace operator and (14), we have  
\begin{equation}
\tr (\mathbf w \mathbf w^H\mathbf  c_k \mathbf c_k^H)=\tr (\mathbf c_k^H \mathbf w \mathbf w^H\mathbf  c_k)=\tr(v_k v_k^H)=|v_k|^2 ,
\end{equation}
which implies that we can rewrite the linear constraint as, 
\begin{equation}
\tr(\mathbf W \mathbf  c_k\mathbf c_k^H)=|v_k|^2.
\end{equation}
The convex LCCMA can now be formulated  as:
\begin{subequations}  
\begin{equation}
\mathbf{\hat  W} =\argmin_{\mathbf {W}} \{( \frac{1}{N}\sum_{n=1}^N  |z(t_{n})-1|^2 )+\tr( \mathbf W)\}, 
\end{equation}
subject to:
\begin{equation}
\tr (\mathbf{W}\mathbf{x}(t_n) \mathbf{x}(t_n))^H)=z(t_n)\;\;\;\;\;\text{n=1,...,N.} \end{equation}
\begin{equation}
\tr (\mathbf{W} \mathbf{c}_k \mathbf{c}_k^H)=|v_k|^2\;\;\;\;\;\text{k=1,...,K.}
\end{equation}
\begin{equation}
\mathbf{W}\succcurlyeq 0.
\end{equation}
\end{subequations}

\begin{figure*}
\centering
\subfigure{
\includegraphics[scale=0.47,trim=0.65cm 0.10cm 0.65cm 0.5cm , clip=true]{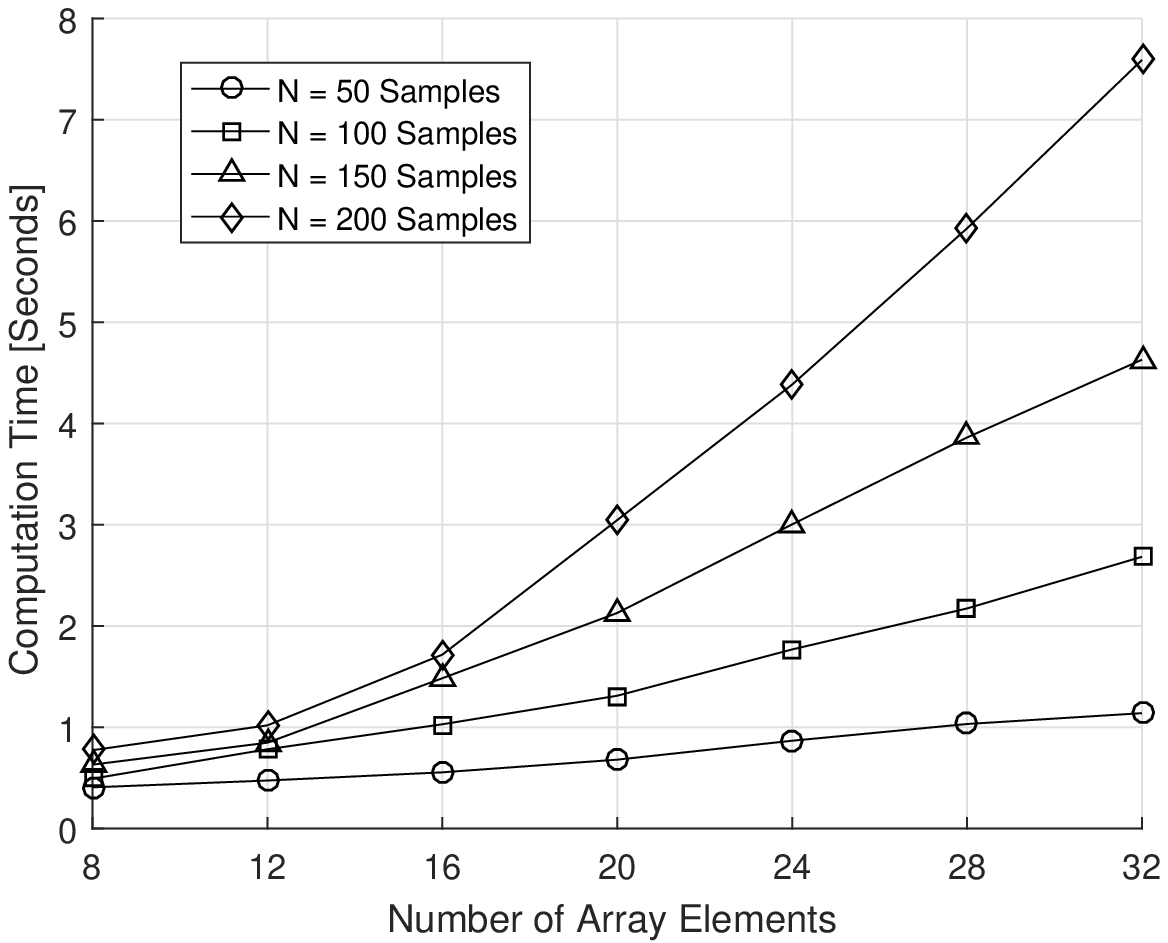}}
\subfigure{
\includegraphics[scale=0.4,trim=0.65cm 0.10cm 0.65cm 0.55cm , clip=true]{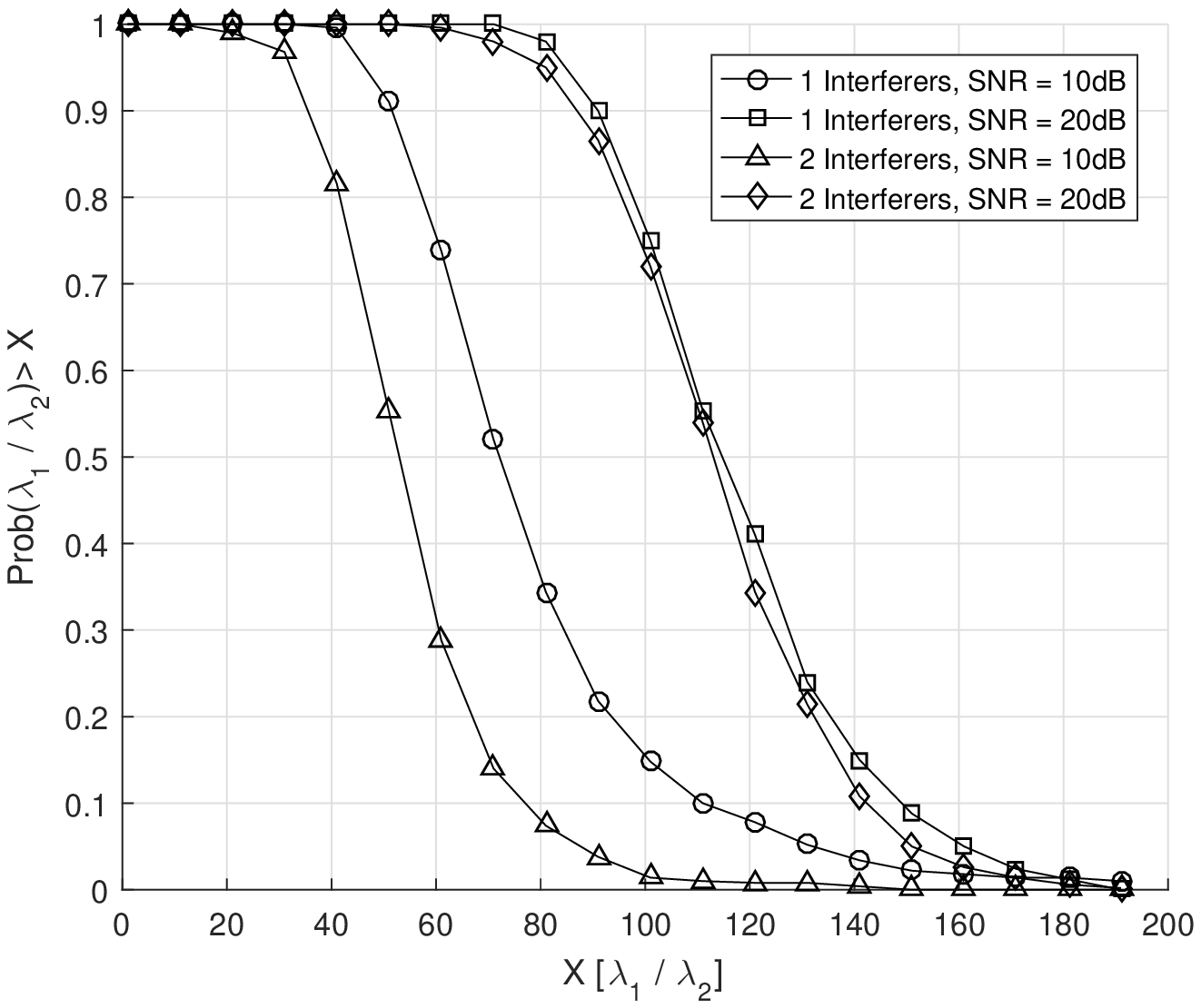}}
\subfigure{
\includegraphics[scale=0.4,trim=0.65cm 0.10cm 0.65cm 0.55cm , clip=true]{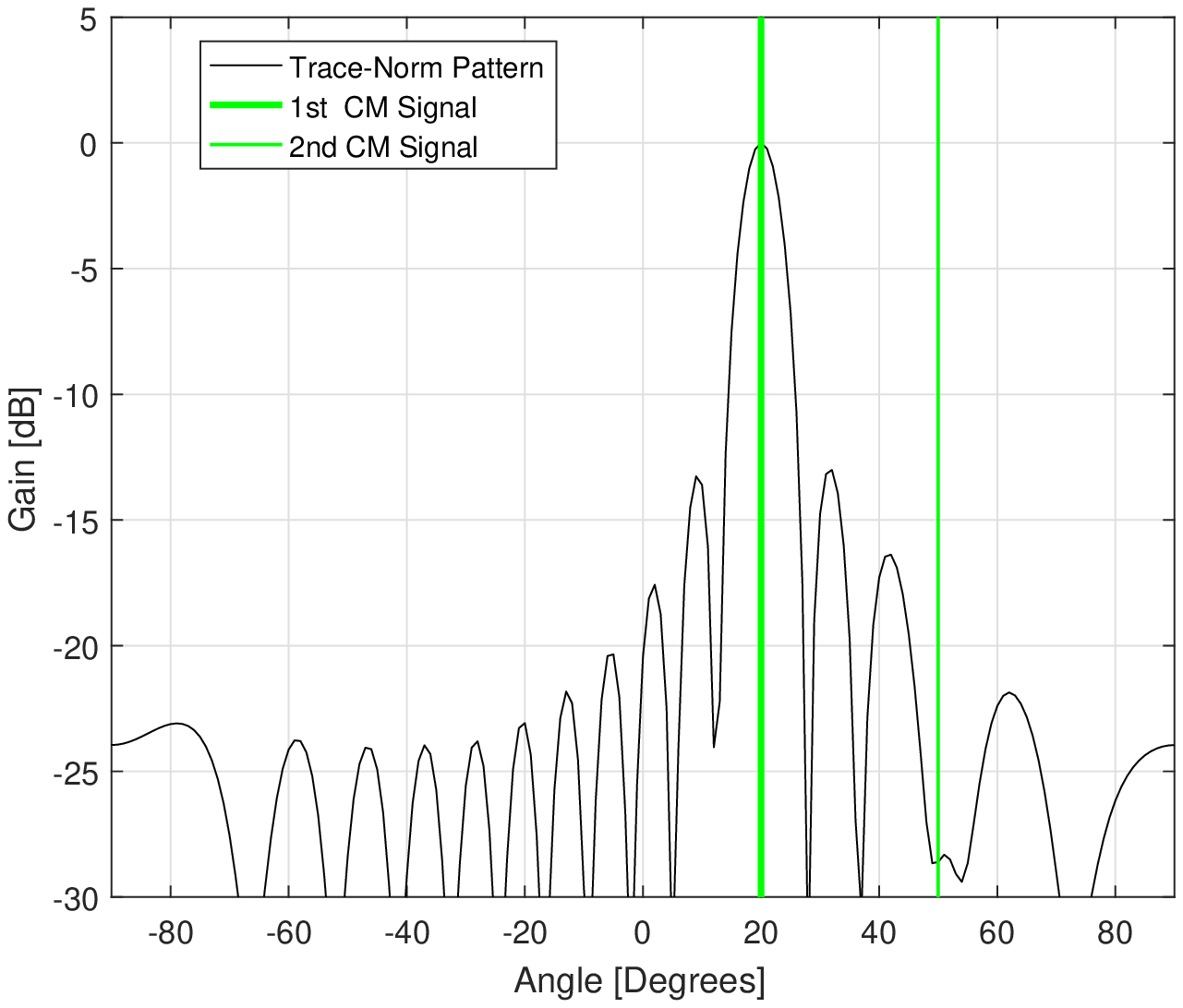}}
\caption{(a) CVX-based computation time of the Trace-Norm approach  vs. number of elements ($P$) and  number of samples ($N$). (b) The ratio between the first and second largest eigenvalues of  $\hat {\mathbf W}$. (c) Averaged array pattern of the Trace-Norm solution, over 1000 experiments, with two CM\ signals: unit power from $20\degree$, and  attenuated by random attenuation ($0dB$ to $-5dB$) from $60\degree$.}
\end{figure*}

\begin{figure*}
\centering
\subfigure{
\includegraphics[scale=0.415,trim=0.75cm 0.10cm 0.65cm 0.7cm, clip=true]{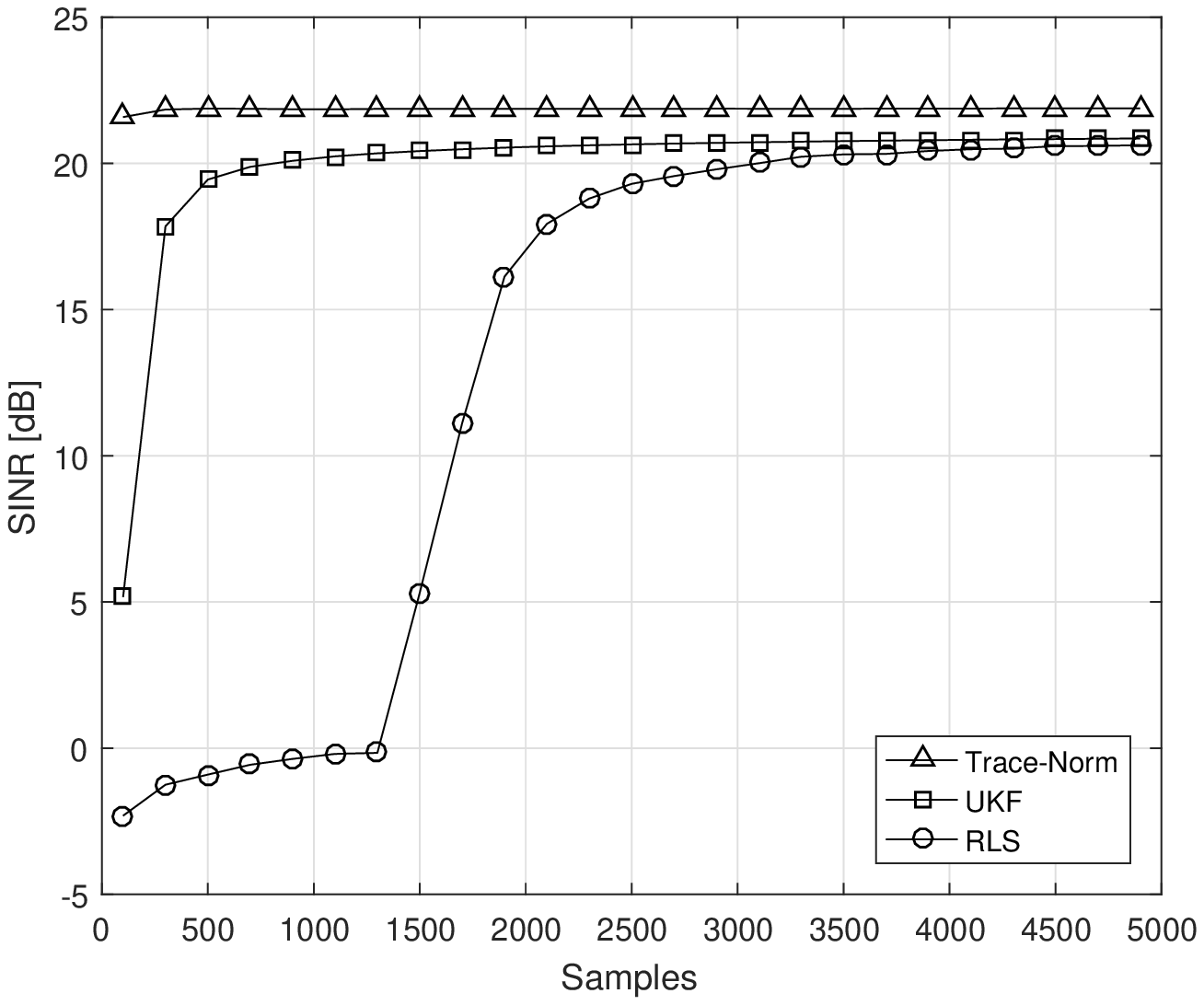}}
\subfigure{
\includegraphics[scale=0.435,trim=0.65cm 0.10cm 0.65cm 0.65cm , clip=true]{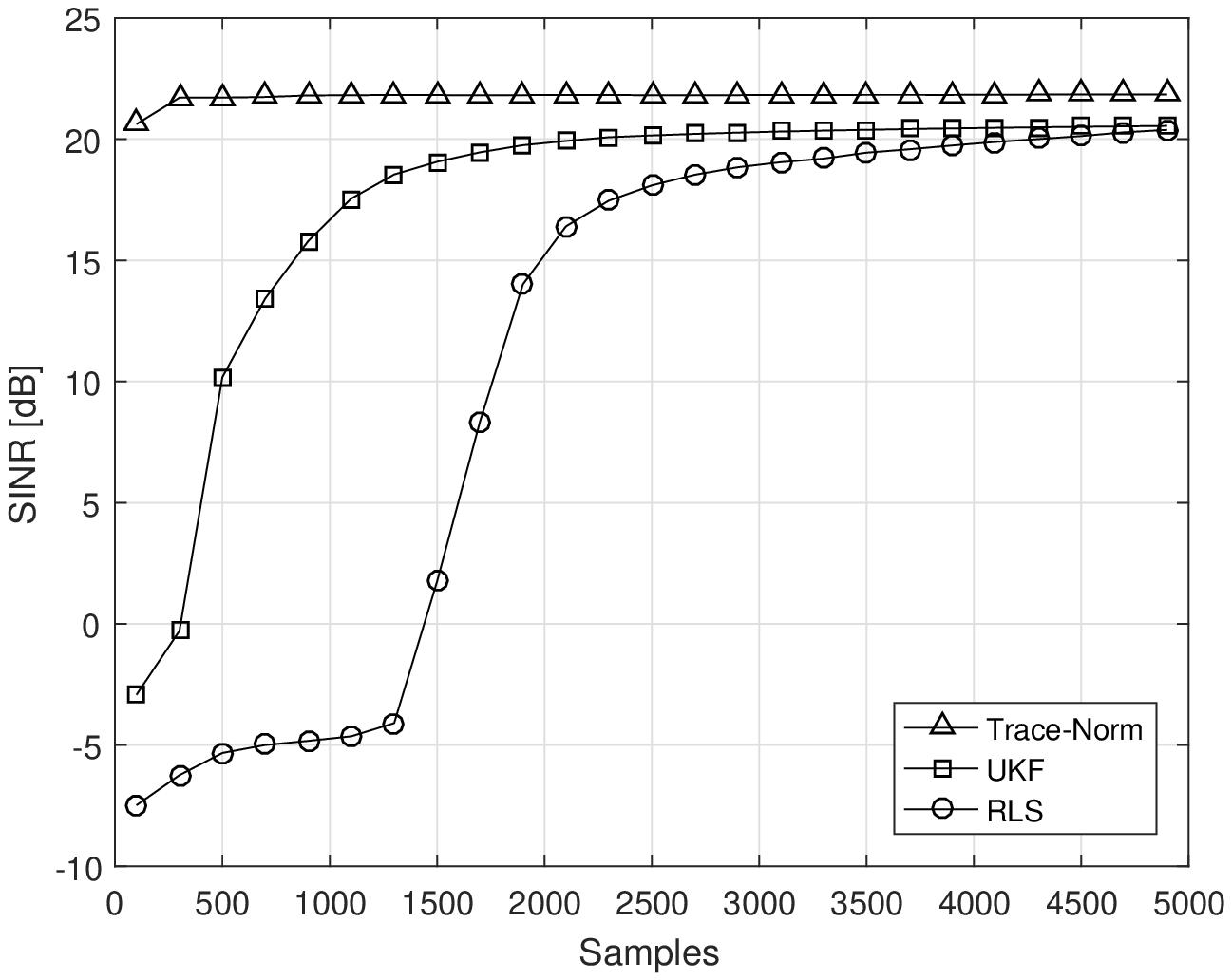}}
\subfigure{
\includegraphics[scale=0.43,trim=0.65cm 0.10cm 0.65cm 0.7cm , clip=true]{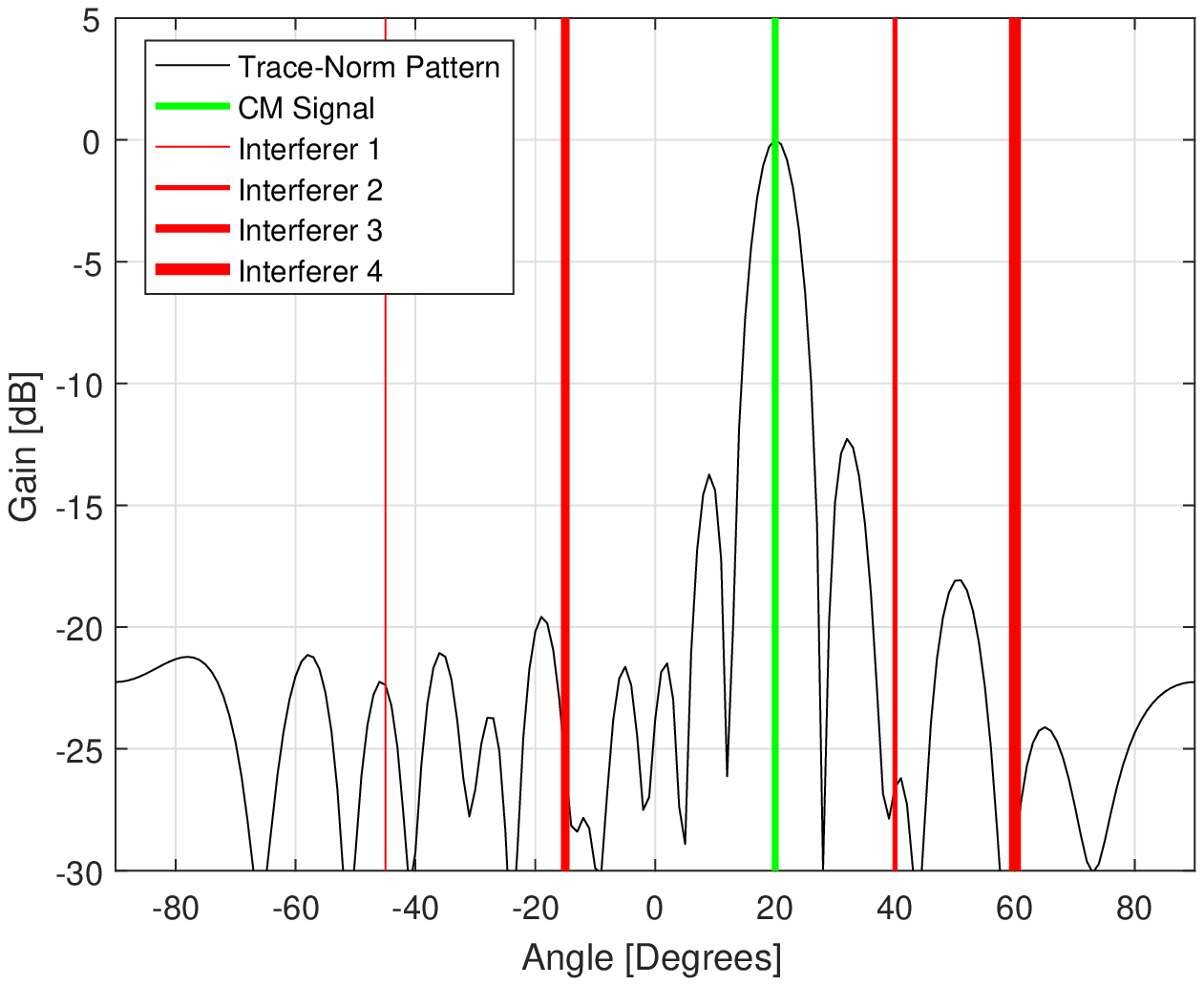}}
\caption{(a) SINR performance of the Trace-Norm, UKF  and RLS ($p=1$,  $\lambda=0.985$  , $\delta=0.001$) vs. number of samples (N), with noise variance $\sigma^{2}_n=0.1$, $P=16$ elements ULA, CM signal at $20\degree$; and three interferers at $-45\degree$,$-15\degree$ and $40\degree$; (b) with four interferers at $-45\degree$,$-15\degree$ , $40\degree$ and $60\degree$.  (c) The resulting array pattern of the Trace-Norm solution, with 4 interferers, avergared over 1000 experiments.}
\end{figure*}
\begin{figure*}
\centering
\subfigure{
\includegraphics[scale=0.43,trim=0.75cm 0.10cm 0.65cm 0.65cm, clip=true]{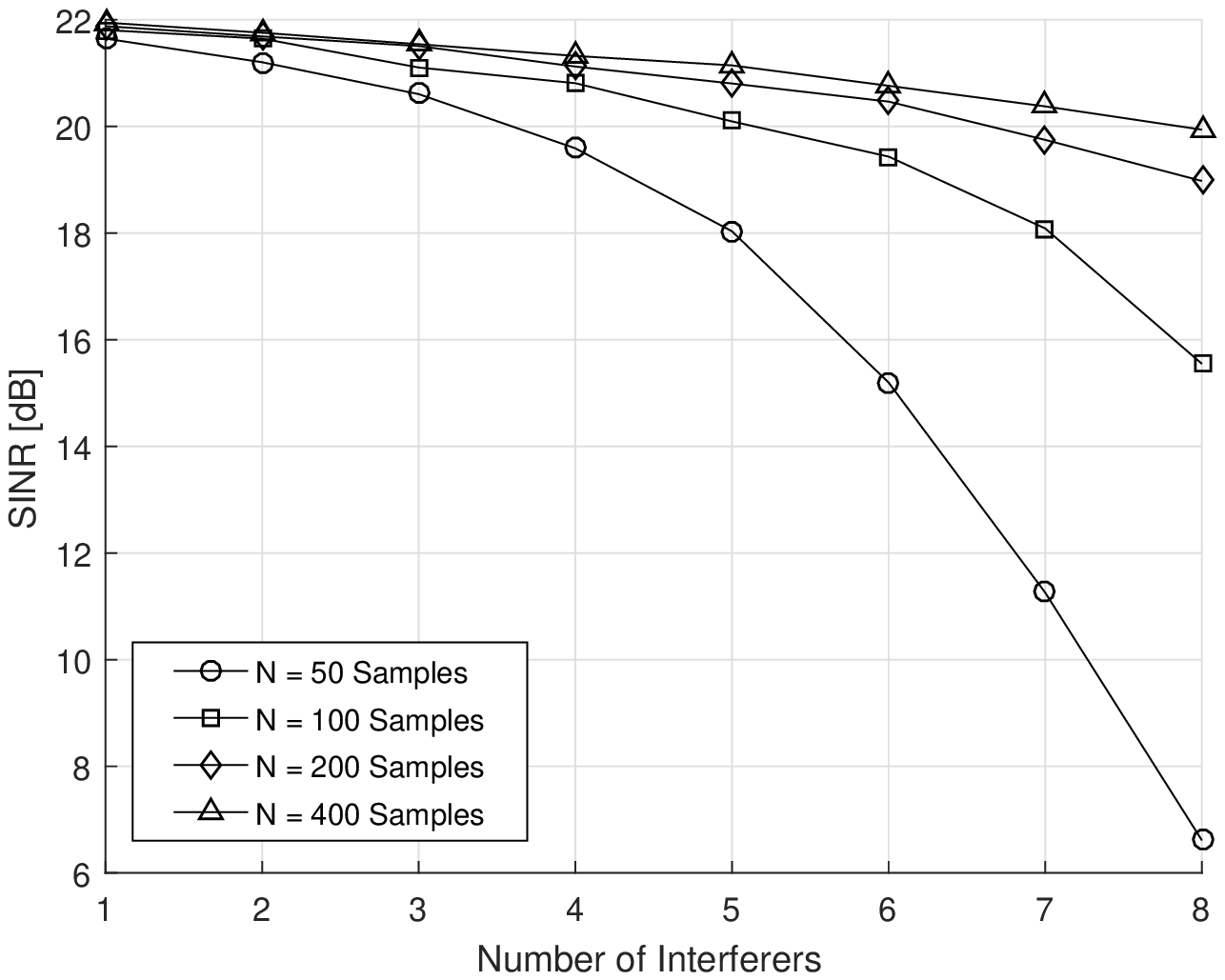}}
\subfigure{
\includegraphics[scale=0.43,trim=0.65cm 0.10cm 0.65cm 0.65cm , clip=true]{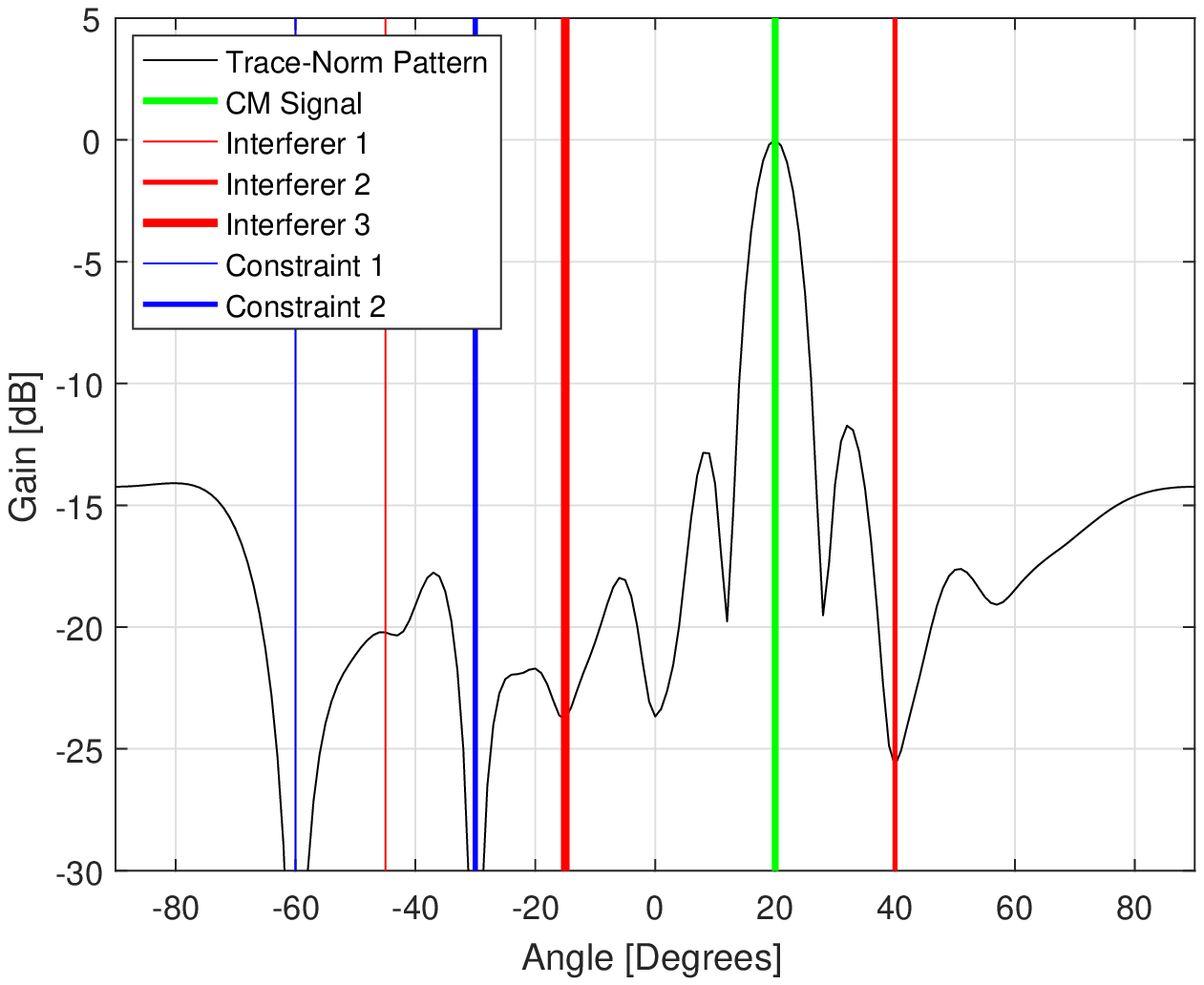}}
\subfigure{
\includegraphics[scale=0.41,trim=0.65cm 0.10cm 0.65cm 0.75cm , clip=true]{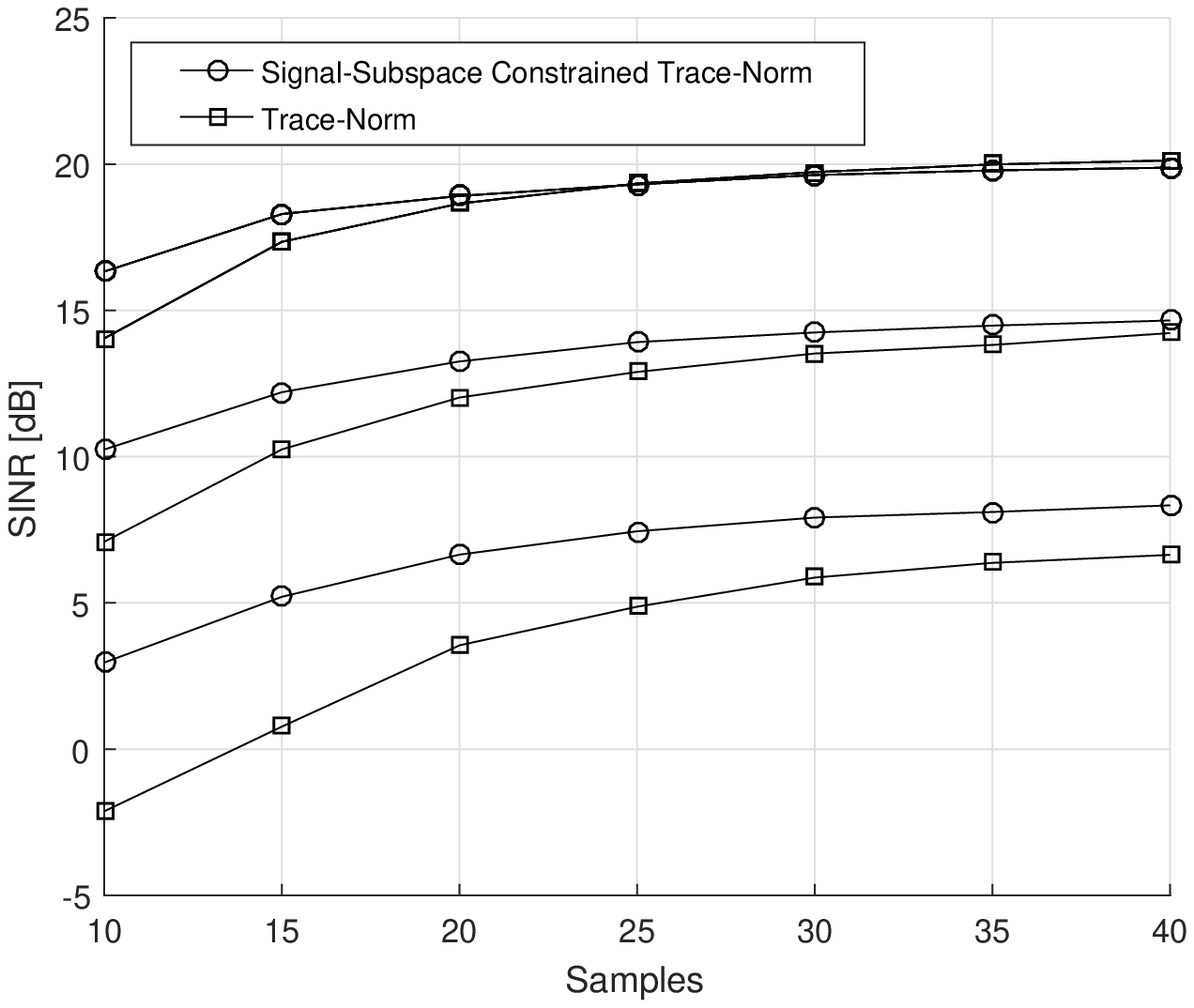}}
\caption{(a) SINR of the Trace-Norm solution vs. the number of interferers (SNR=10dB, P=16). (b)  LCCMA Trace-Norm array pattern, with null constraints at $-30\degree$ and at $-60\degree$, and three interferers ($N = 200$  samples, $\sigma^{2}_n=0.1$, $P=16$ elements). (c) SINR results of the Trace-Norm approach vs. the Signal-Subspace Constrained LCCMA Trace-Norm ($P=32$ elements) in the presence of a CM signal at $20\degree$, and  2 interferers at $-45\degree$,$-20\degree$:  SNR = -5dB (lower curves); SNR = 0dB (middle curves); and  SNR = 5dB (upper curves).}
\end{figure*} 
\section{Performance Evaluation}
In this section we present computation time and simulation results illustrating the  performance of our solution, referred to as Trace Norm. The performance is compared to the Recursive Least Squares (RLS) [8] and the Unscented Kalman Filter (UKF) [9]  solutions.  

The desired signal was simulated as a unit power QPSK signal. The interfering signals were simulated  as complex  Gaussian with zero mean and unit variance. The noise was simulated as a complex Gaussian with zero mean and covariance $\sigma^{2}_{n}\mathbf I $. The performance measure employed is the signal-to-interference-plus-noise  ratio (SINR) at the beamformer output:
\begin{equation}
SINR=\frac{\mathbf w^{H}R_{ss}\mathbf w}{\mathbf w^{H}R_{nn}\mathbf w+\mathbf w^{H}R_{ii}\mathbf w},
\end{equation}
where $R_{ss}= \mathbf a(\theta_{0}) \mathbf a(\theta_{0})^{H}$, $R_{nn}=\sigma^{2}_{n}\mathbf I$, and $R_{ii}=\sum_{j=1}^{q}\mathbf a(\theta_{j}) \mathbf a(\theta_{j})^{H} $are the CM signal, noise and interference covariance matrices, respectively. All presented results are averaged over 100  experiments, unless specified differently.

\indent Experiment 1 evaluates the computation time of the Trace Norm solution.  The worst case computational complexity of a general convex optimization problem is given  by $O(\max(P,N)^4 N^{0.5} \log(1/\epsilon))$ [17],
where $\epsilon$ is the solution accuracy. To provide more typical results we  evaluated the computation time using the MATLAB-based CVX [20]   toolbox, and the results are presented in Fig. 1(a)  Note that  the speed-up factor between CVX-based implementation and a real-time implementation, as  analyzed  in [21], is in the range of $\times 100$ to $\times 10,000 $ (single processor).   The simulated\footnote{Using an Intel Core i7-5930K, 32GB RAM, desktop computer.} scenario includes a    CM signal impinging from  $20\degree$ on a Uniform Linear Array (ULA) with $P=8$ to $P=32$ elements, and 3 interferers impinging from $-45\degree$ , $-15\degree$ and $40\degree$  (noise variance $\sigma^{2}_n=0.1)$. 
\indent

Experiment 2 evaluates the ratio between the largest ($\lambda_1$) and the second largest ($\lambda_2$)  eigenvalues of  $\hat {\mathbf W}$, which is a good measure for the goodness of the rank-1 approximation of the trace norm solution of  $\hat {\mathbf W}$. We evaluated this ratio by solving 500 times\footnote{Each solution treadted different transmitted symbols, different noise realization, and different interfering signals waveforms.} each of the following scenarios: a CM signal in the presence of 0,1, or 2 interferers, all signals are of equal power, at SNR of 10dB or 20dB ($\sigma^{2}_n=0.1$ or $0.01$, respectively). For the case of no interference, the ratio $\frac{\lambda_1}{\lambda_2}$ exceeded $10^6$ with probability 1, implying a perfect rank-1 result. Fig. 1(b) presents the results for the cases of 1 and 2 interferers, and reveals that $\frac{\lambda_1}{\lambda_2}\ge 10$, with probability 1, for SNR = 10dB, and$\frac{\lambda_1}{\lambda_2}\ge 50$ for SNR = 20dB. These results demonstrate the goodness of the rank-1 approximation of $\hat {\mathbf W}$.
\indent

Experiment 3 evaluates the performance of the Trace-Norm solution in the presence of two CM signals: The first from $20\degree$ with unit power,  and the second from $50\degree$ ,attenuated in each trial by a random attenuation, uniformly distributed between 0dB to -5dB. Fig. 1(c) presents the averaged array pattern, over 500 experiments, and demonstrates the "capture" effect of the Trace-Norm solution: the algorithm captures always the strongest CM signal, and cancels the weaker. 

Experiment 4 compares the SINR of the Trace-Norm, UKF and  RLS, in the presence of interferers. Note that since the Trace-Norm is a batch approach, whereas UKF and RLS are on-line approaches (processing one sample at time), the reported SINR, at each sample index $n$ , means that the algorithm processed all samples from the 1st until the $n$-th.   In the first scenario we  simulated a  CM\ signal  impinging  from  $20\degree$ on a 16 elements ULA, with 3 interfering signals impinging from $-45\degree$ , $-15\degree$ ,$40\degree$, and noise variance $\sigma^{2}_n=0.1$. The results are presented in Fig. 2(a) and demonstrate that the Trace-Norm solution obtain better SINR with only 100 samples, whereas UKF\ converges after $N=500$ samples, and RLS after $N=2,700$ samples. Fig. 2.(b). presents the performance with an additional interferer from 60\degree. In this case convergence of the UKF and RLS\ is slower ($N=1,500$ and $N=3,500$ samples, respectively), whereas the Trace-Norm is essentially invariant to to the addition of the interferer, and surpasses  UKF\ and RLS with only 100 samples. The array pattern of the Trace Norm with $N=200$ samples (averaged over 1,000 experiments), is depicted in Fig. 2(c). The  rejection of all 4 interferers is clearly visible.
\indent

Experiment 5 presents the SINR of the Trace-Norm solution vs. the number of interferers. The simulated scenario includes   a  CM\ signal  impinging  from  $-25\degree$ on a 16 elements ULA, with a varying number of interferers between 1 to 8, impinging from directions chosen randomly from the following set of directions: $-85\degree, -70\degree, -55\degree, -40\degree,-10\degree ,5\degree, 20\degree,35\degree,50\degree,65\degree$ and $80\degree$. The noise variance per array element is $\sigma^{2}_n=0.1$, corresponding to SNR=10dB for all signals. The results presented in Fig. 3(a), demonstrate that the Trace-Norm solution\ can handle effectively (providing SINR\textgreater20dB)\ 3 interferers with $N=50$ samples, and 7 interferers with $N=400$ samples.

Experiment 6 demonstrates the ability of the Trace-Norm LCCMA to generate deep nulls in the array pattern  in predefined directions, using the constraint  (11),(12). The simulated scenario includes a CM signal at $20\degree$ impinging on a 16 element array, and 3 interferers from  $-45\degree$,$-15\degree$ and $40\degree$ ($\sigma^{2}_n=0.1$). The nulls are constrained to  directions $-30\degree$ and $-60\degree$. The resulting array pattern, averaged over 1,000 experiments,  is depicted in Fig. 3(b). Clearly visible is  the rejection of all interferers, as well as the deep nulls in the specified directions.
\indent

Experiment 7 demonstrates the performance advantage of the  Trace Norm LCCMA  over  the Trace-Norm CMA when the constraint (11),(13) is imposed. The  simulated scenario includes   a CM\ signal  impinging  from  $20\degree$ on a 32 elements ULA, with 2  interferers impinging from $-45\degree$ and $-20\degree.$ The SNR   per array element is varied between -5dB to 5dB.  The constraint (11),(13) forces the beamforming  vector to be confined  to the \textit{3-dimensional signal subspace}. Fig. 3(c) shows SINR results vs. the number of  samples (N). The results demonstrate the advantage of the Trace Norm LCCMA over the Trace-Norm CMA for  all signal-to-noise ratios (excluding a minor disadvantage for SNR=5dB and N\textgreater30 samples).

\section{Conclusions}
We have presented  new convex-optimization-based solutions for the CMA and for the related problem of LCCMA.
Our CMA solution was shown to provide much better  performance than existing solutions based on UKF and RLS. Moreover, the SINR of our solution, was shown to approach the theoretical limit even for relatively small  number of samples.   
We have also shown that our  LCCMA solution  enables the incorporation of a variety of linear constraints on the beamformer vector in a simple and effective way. We have shown that apart from enabling unity gain and null constraints to predefined directions, we can also incorporate more general constraints such as constraining the beamformer vector to the signal subspace. This was shown to provide significant performance gain as compared to unconstrained CMA. \

\appendices  



\ifCLASSOPTIONcaptionsoff
  \newpage
\fi


%

\end{document}